\documentclass[reprint,superscriptaddress,
amsmath, amssymb,aps, prd,notitlepage,longbibliography,floatfix,
nofootinbib 
]{revtex4-1}

\usepackage{tensor}     
\usepackage{graphicx}   
\usepackage{subfigure}
\usepackage[
colorlinks=true,        
citecolor=blue,         
linkcolor=blue,         
urlcolor=blue           
]{hyperref}             
\usepackage{bm}         
\usepackage{xcolor}     
\usepackage{lipsum}
\usepackage{color}      
\usepackage[utf8]{inputenc} 
\usepackage[section]{placeins} 
\usepackage{multirow}
\usepackage[title]{appendix}
\usepackage{simpler-wick}
\usepackage{float}
\newcommand{\nc}{\newcommand*}

\nc{\al}{\alpha}
\nc{\s}{\sigma}
\nc{\kp}{\kappa}
\nc{\dt}{\delta}
\nc{\Dt}{\Delta}
\nc{\Ld}{\Lambda}
\nc{\p}{\partial}
\nc{\Gm}{\Gamma}
\nc{\om}{\omega}
\nc{\Om}{\Omega}
\nc{\rd}{\mathrm{d}}
\def\({\left(}
\def\){\right)}
\def\[{\left[}
\def\]{\right]}
\def\e{\begin{equation}}
\def\q{\end{equation}}
\def\be{\begin{equation}}
\def\ee{\end{equation}}
\def\m{\begin{eqnarray}}
\def\n{\end{eqnarray}}
\def\beq{\begin{eqnarray}}
\def\eeq{\end{eqnarray}}
\nc{\Eq}[1]{Eq.~\eqref{#1}}     
\nc{\Fig}[1]{Fig.~\ref{#1}}     
\nc{\Table}[1]{Table~\ref{#1}}  
\nc{\Sec}[1]{Sec.~\ref{#1}}     
\nc{\Msun}{M_\odot}             
\nc{\Ogw}{\Omega_{\mathrm{GW}}}
\nc{\gpcyr}{\mathrm{Gpc}^{-3}\,\mathrm{yr}^{-1}}
\nc{\lvc}{LIGO/Virgo} 
\nc{\SNR}{\mathrm{SNR}} 
\nc{\rhoGW}{\rho_{\mathrm{GW}}}
\nc{\vd}{\vec{d}}
\nc{\av}[1]{\langle #1 \rangle} 
\nc{\km}{\mathrm{km}}
\nc{\Mpc}{\mathrm{Mpc}}
\nc{\Tobs}{T_{\mathrm{obs}}}
\nc{\fyr}{f_{\mathrm{yr}}}

\nc{\addref}{[\textcolor{red}{add ref}] } 
\nc{\eg}{\textit{e.g.~}}
\nc{\app}{\approx}
\nc{\hf}{\frac{1}{2}}
\nc{\discuss}{\textcolor{red}{Add discussion here!}}
\nc{\red}[1]{\textcolor{red}{#1}}
\nc{\hp}{h_+} 
\nc{\hc}{h_{\times}} 
\nc{\mbh}{M_{\rm BH}}
\nc{\mdisk}{M_{\mathrm{disk}}}
\nc{\rdisk}{r_{\mathrm{disk}}}
\nc{\mc}{M_{\mathrm{c}}}

\begin{document}
	
\title{Gravitational Waves from Accretion Disks: \\Turbulence, Mode Excitation and Prospects for Future Detectors}
	

\author{Chen Yuan}
\email{chenyuan@tecnico.ulisboa.pt}
\affiliation{CENTRA, Departamento de Física, Instituto Superior Técnico – IST, Universidade de Lisboa – UL, Avenida Rovisco Pais 1, 1049–001 Lisboa, Portugal}

\author{Vitor Cardoso} 
\affiliation{Niels Bohr International Academy, Niels Bohr Institute, Blegdamsvej 17, 2100 Copenhagen, Denmark}
\affiliation{CENTRA, Departamento de F\'{\i}sica, Instituto Superior T\'ecnico -- IST, Universidade de Lisboa -- UL, Avenida Rovisco Pais 1, 1049-001 Lisboa, Portugal}

\author{Francisco Duque} 
\affiliation{Max Planck Institute for Gravitational Physics (Albert Einstein Institute) Am Mühlenberg 1, D-14476 Potsdam, Germany}

\author{Ziri Younsi} 
\affiliation{Mullard Space Science Laboratory, University College London, Holmbury St.~Mary, Dorking, Surrey, RH5 6NT, United Kingdom.}


\date{\today}

\begin{abstract}
We study gravitational-wave emission by turbulent flows in accretion disks around spinning black holes or neutron stars. We aim to understand how turbulence can stochastically excite black hole quasinormal ringing and contribute to a stochastic gravitational-wave background from accretion disks around compact objects. We employ general relativistic magnetohydrodynamic simulations and feed them as the source of the Teukolsky master equation to evaluate the gravitational wave energy spectrum of a single source.
The stochastic gravitational wave background from accretion disks generated by the population of stellar-mass compact objects is far below the sensitivity of third-generation ground-based detectors. In contrast, the supermassive black hole population, in particular those actively accreting, could lead to $\Omega_{\mathrm{GW}}\sim 10^{-15}$ in the microHertz. This signal remains well below the sensitivities of pulsar-timing-arrays and LISA, making direct observation infeasible.
\end{abstract}

\maketitle

\maketitle
\noindent {\bf \em Introduction.} 
The remarkable detection of gravitational waves (GWs) from the merger of two black holes (BHs) by the LIGO-Virgo Scientific Collaboration~\cite{LIGOScientific:2016dsl,LIGOScientific:2021djp} inaugurated a new era in physics~\cite{Yunes:2013dva,Berti:2015itd,Barack:2018yly,Cardoso:2019rvt}. The understanding of mechanisms for BH formation and growth across cosmic time \cite{Berti:2015itd,Barack:2018yly}, of possible phase transitions in the early universe \cite{Witten:1984rs,Kosowsky:1992rz,PhysRevD.45.4514,Kamionkowski:1993fg,NANOGrav:2021flc,Xue:2021gyq,Romero:2021kby,Jiang:2022mzt}, or the nature of dark matter are now within reach \cite{Sasaki:2016jop,Yuan:2021qgz,Chen:2019xse,Yuan:2022bem,Tsukada:2018mbp,Tsukada:2020lgt,Yuan:2021ebu,Preskill:1982cy,Abbott:1982af,Dine:1982ah,Arvanitaki:2009fg,Arvanitaki:2010sy,Essig:2013lka,Brito:2015oca,Marsh:2015xka,Hui:2016ltb,Annulli:2020lyc,Chadha-Day:2021szb}. Even more intriguing is the opportunity to test gravity in the strong-field regime, particularly near BH horizons. Of particular interest to us here are rotating BHs surrounded by an accretion disk.

Accretion disks are interesting probes of strong gravitational fields and of extreme astrophysical processes, where the role of BH rotation, surrounding plasma and magnetic fields is yet to be fully understood. The dynamics of accretion disks are incredibly complex and rich. Among others, one of the crucial factors is turbulence~\cite{Balbus:1991ay,Balbus:1998ja}, which breaks the axial symmetry of the system and leads to the emission of light but also GWs, providing new avenues for understanding the behaviour of accretion disks. In the same way that turbulence in the Sun drives the stochastic excitation of its characteristic modes~\cite{1977ApJ...211..934G,1977ApJ...212..243G,Christensen-Dalsgaard:2002ney}, it might be expected that turbulent accretion can excite modes of BHs, but a precise calculation has not been done. An order of magnitude estimate was attempted in the past~\cite{Araya-Gochez:2003hzq}, but missed all the intricate features of astrophysical disks. The formation mechanism of collapsar disks was recently studied \cite{Gottlieb:2024dfw}, with the conclusion that the resulting GW emission could reach the sensitivities of current and future ground-based detectors. This -- the possible stochastic excitation of quasinormal modes of BHs -- was, in fact, our original motivation to understand GW emission from systems including a BH and an accretion disk.

The superposition of GW sources will form a stochastic GW background (SGWB). The study of the SGWB generated by accretion disks is instrumental in refining our theoretical models of accretion physics and stellar populations. As far as we know, there is little to no work studying the SGWB generated by accretion disks. In this Letter, we address this problem.
We adopt geometrical units $G=c=1$.

\noindent {\bf \em Numerical setup of accretion disk.} 
\begin{figure*}[htbp!]
\centering
\includegraphics[width=0.45\textwidth]{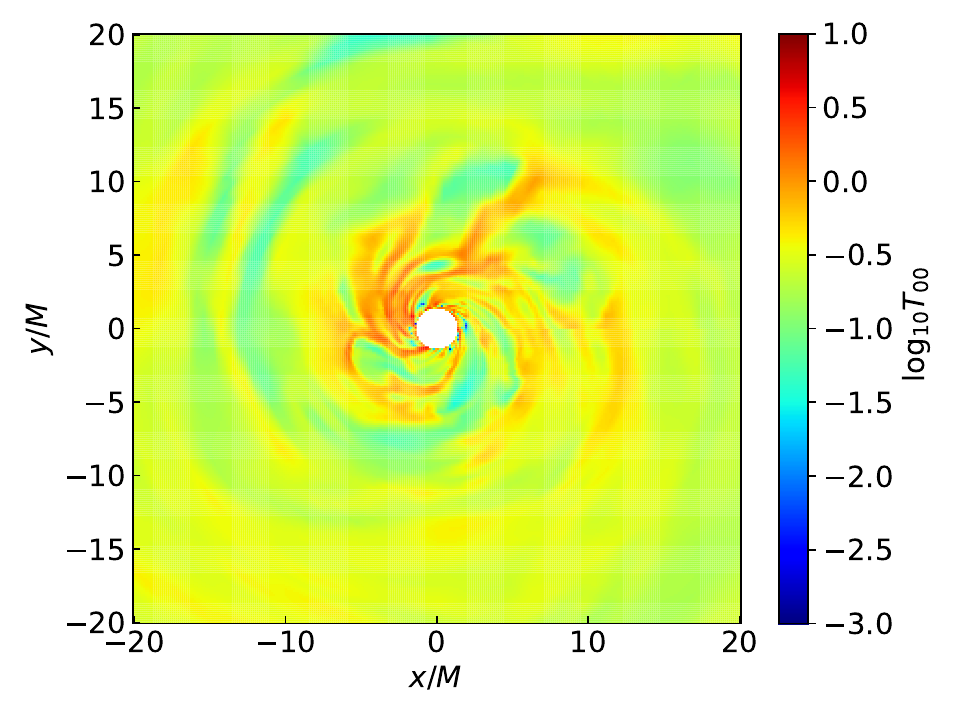}
\includegraphics[width =0.45\textwidth]{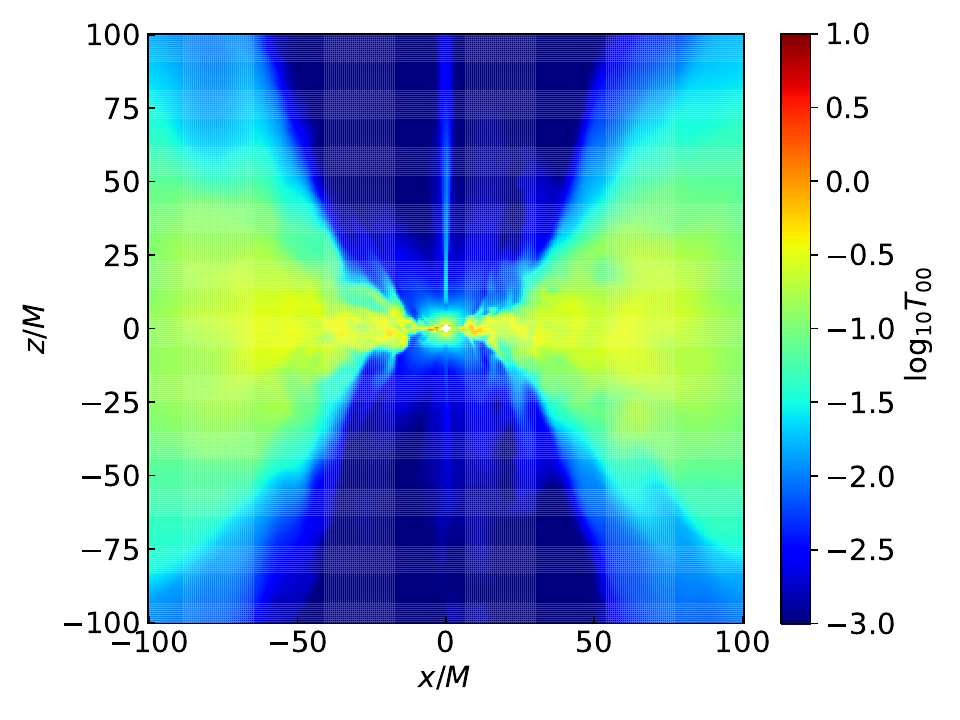}
\caption{\label{density} Initial snapshots of the rest mass density ($T_{00}\equiv\rho$) of the accretion disk in our simulation, before rescaling.
\textit{Left panel}: Cross-sectional cut through the equatorial plane of the disk ($z=0$).
\textit{Right panel}: Two-dimensional cross-section of the simulation in the $x$--$z$ plane, showing the extended, turbulent disk and low-density funnel regions both above and below the middle plane comprising the jets.
\label{snapshot_disk}
}
\end{figure*}
We consider a setup where a BH of mass $M$, and dimensionless angular momentum $\chi$, is surrounded by an accretion disk. 
We assume the disk back-reacts weakly on the background spacetime, such that for the purposes of matter evolution the spacetime may be assumed to be stationary and described by the Kerr metric.
The simulations of the matter around the BH are computed using the \texttt{BHAC} code~\cite{Porth:2016rfi,Olivares:2019dsc}.

We focus on a Kerr BH with $\chi=0.94$, adopting spherical Kerr-Schild coordinates with a logarithmic radial coordinate, i.e., $x^{1}=\ln r$, focusing resolution in regions closer to the event horizon.
The simulation outer boundary is placed at $r=2500~M$ and the inner boundary is well within the event horizon.
The grid resolution is $384\times 192 \times 192$ in $x^{i}$ ($i=1,\, 2, \, 3$).
The GRMHD simulation is initialised with a Fishbone-Moncrief hydrodynamic equilibrium torus profile~\citep{FM1976}, with $r_{\rm in}=20\,M$ and $r_{\rm max}=40\,M$.
We employ an ideal gas equation of state with relativistic adiabatic index $\Gamma=4/3$.
The equilibrium torus profile is then suffused with a single weak magnetic field loop, with the radial distribution of the magnetic field profile chosen to ensure sufficient magnetic flux is deposited onto the BH, enabling the
magnetically arrested disc state to be reached \citep{Narayan2003,Tchekhovskoy2011}.
The magneto-rotational instability inside the torus is triggered by applying 2 per cent of a random perturbation to the torus gas pressure. 
The treatment of very low-density regions and regions of high magnetisation is performed in the conventional manner for GRMHD simulations of BHs as follows \citep[e.g.,][]{Mizuno2021}.
Floor values are applied to the rest-mass density as $\rho_{\rm floor}=10^{-4} \, r^{-2}$ and to the gas pressure as $P_{\rm floor}=\left(10^{-6}/3\right) \, r^{-2 \Gamma}$.
These ensure that in all grid cells where $\rho \le \rho_{\rm floor}$ or $P \le P_{\rm floor}$, one sets $\rho = \rho_{\rm floor}$ and $P = P_{\rm floor}$.
We also introduce a ceiling within regions of high
magnetisation, $\sigma$, such that $\sigma_{\rm max}=100$ for all grid cells where $\sigma \ge \sigma_{\rm max}$.

In order to evaluate physical quantities, we begin with the stress-energy tensor of a magnetised perfect fluid:
\begin{equation}
    T_{\mu\nu} = (\rho + P + \rho\epsilon + b^2)u_{\mu}u_{\nu} + \left(P + \frac{1}{2}b^2\right)g_{\mu\nu} - b_{\mu}b_{\nu} \,,
\end{equation}
where $\rho$ and $P$ represent, respectively, the density and pressure of the fluid, $\epsilon$ is the internal energy density, $g_{\mu\nu}$ denotes the metric tensor, $u_{\mu}$ is the four-velocity of the fluid, and $b^2 := b_{\mu} b^{\mu}$, wherein $b_{\mu}$ is the magnetic field four-vector.
We work in Lorentz-Heaviside units, absorbing a factor of $\sqrt{4\pi}$ into the definition of $b_{\mu}$.

After evolving the system for $10^4~M$ to reach a quasi-stationary state, we extract $T_{\mu\nu}$ with a time step of $\Delta t = 10~M$ for a duration of $2000~M$.
In order to circumvent the high memory and storage requirements, we uniformly down-sample the GRMHD data by a factor of two ($166\times 96 \times 96$) when computing $T_{\mu\nu}$, further restricting ourselves to the domain $r_{+} < r \le 1000~M$, where $r_{+}:=M+\sqrt{M^{2} - a^{2}}$ is the event horizon radius of the BH.
An initial snapshot of the energy density is shown in Fig.~\ref{snapshot_disk}, where turbulent-like features are apparent (spiral structures and uneven energy density distribution along the polar directions, indicating the presence of hydrodynamic turbulence).

To evaluate the SGWB from accretion disks, one needs to consider disks with different masses and mass accretion rates.
These may be obtained by re-scaling the simulation code data, $T_{\mu\nu}^{(\mathrm{code})}$, to physical CGS units, as done in radiative transfer post-processing \citep{Younsi2012,Younsi2023}.
The simulation code density scales as $\rho^{(\mathrm{cgs})}= \rho^{\rm unit}~\rho^{(\mathrm{code})}$.
Introducing the gravitational radius $r_{\rm g}\equiv GM/c^{2}$, here $\rho^{\rm unit} \equiv \mathcal{M}/r_{\rm g}^{3}$ and $\mathcal{M} := \dot{M} \left(r_{\rm g}/c\right)$ is a physical re-scaling factor set by the BH's mass accretion rate ($\dot{M}$) and its mass.
The stress-energy tensor of the fluid then scales to physical units by a factor of $c^{2}\rho^{\rm unit}$
\begin{equation}
    T_{\mu\nu}^{(\mathrm{cgs})} = \left(\frac{\mathcal{M} c^{2}}{r_{\rm g}^{3}}\right) \, T_{\mu\nu}^{(\mathrm{code})} \,.
\end{equation}
We parameterize the accretion rate of the central object as
\begin{equation}
\dot{M} = f_{\mathrm{Edd}} \dot{M}_\mathrm{Edd} \, ,
\end{equation}
where the Eddington ratio $f_{\mathrm{Edd}}$ is a free parameter that characterizes the accretion rate of the central object in units of the Eddington accretion rate $\dot{M}_\mathrm{Edd}$~\cite{Yuan:2014gma}
\begin{equation}
    \dot{M}_\mathrm{Edd} \simeq 2\times 10^{-8} \(\frac{M}{M_{\odot}}\) M_{\odot}\, \mathrm{yr}^{-1} \, .
\end{equation}
%

\noindent {\bf \em Gravitational radiation.} 
Once we have the time-varying stress-energy tensor of the disk, GW emission can be computed by solving Teukolsky's master equation~\cite{Teukolsky:1972my,Teukolsky:1973ha}. In this approach, the radiative degrees of freedom of the gravitational field at infinity are encoded in a master variable $\Psi$ related to the Newman-Penrose scalar $\Psi_4$ via $\Psi = (r-ia\cos\theta)^4 \Psi_{4}$ and $\Psi_4$ is directly related to the GW polarisations via
\begin{equation}
\Psi_4=\frac{1}{2}\left(\frac{\partial^2 h_{+}}{\partial t^2}-i \frac{\partial^2 h_{\times}}{\partial t^2}\right)\,.
\end{equation}
The Teukolsky master equation is shown in Appendix.~\ref{source}.
%
%
We decompose in azimuthal components $\Phi_m$
\begin{equation}
\Psi(t, r_{*}, \theta, \tilde{\phi})=\sum_{m=-\infty}^\infty r^{3}e^{i m \tilde{\phi}} \Phi_m \left(t, r^{*}, \theta\right) \, .
\end{equation}
Here, $r_*$ and $\tilde{\phi}$ are the tortoise coordinates, which are related to Boyer-Lindquist coordinates through
\begin{equation}
    \begin{aligned}
r^{*} & =  r+\frac{2 M r_{+}}{r_{+}-r_{-}} \ln \frac{r-r_{+}}{2 M}  -\frac{2 M r_{-}}{r_{+}-r_{-}} \ln \frac{r-r_{-}}{2 M}, \\
\tilde{\phi} & =  \phi+\frac{a}{r_{+}-r_{-}} \ln \frac{r-r_{+}}{r-r_{-}},
\end{aligned}
\end{equation}
with $r_\pm=M\pm\sqrt{M^2-a^2}$. The source term $T$ in the Teukolsky euation is also decomposed into $T_m$ components.
We solve the Teukolsky equation numerically for each $m$ mode in the time domain using the two-step Lax-Wendroff method described in \cite{Krivan:1997hc,Lopez-Aleman:2003sib,Pazos-Avalos:2004uyd,Sundararajan:2007jg, Cardoso:2021vjq, Cardoso:2021sip} with second-order finite differences.

\begin{figure}[t]
\centering
\includegraphics[width = 0.45\textwidth]{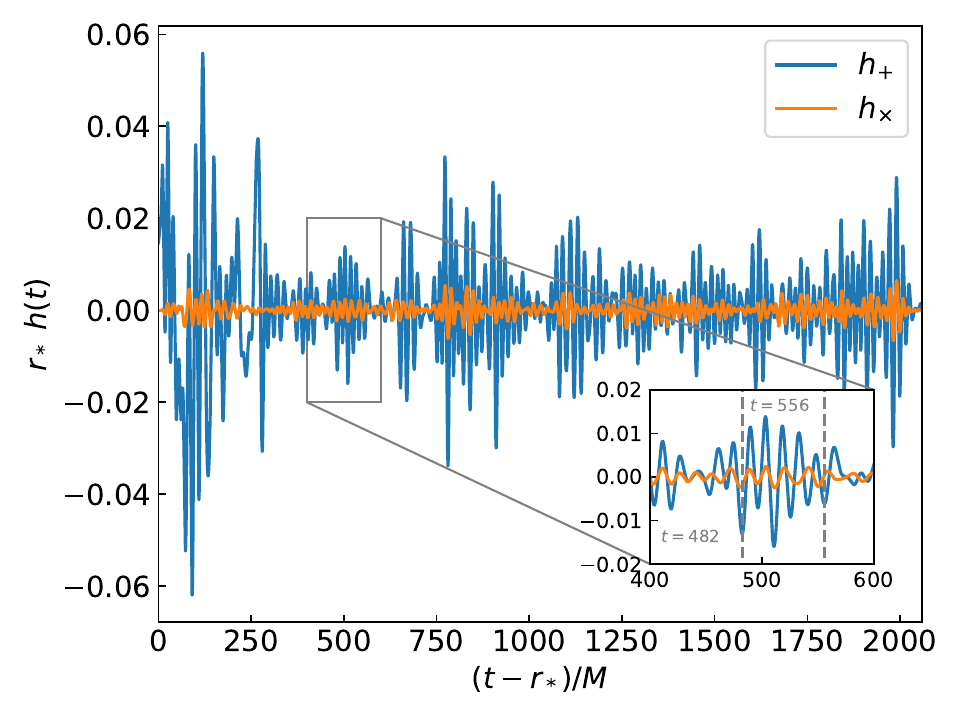}
\caption{The $m=0$ gravitational waveform generated by the accretion disk. The signal is extracted at $r_{*}^{\rm ext}=500M$. From the time-domain data we estimate the frequency of the signal to be of order $M\omega \sim 0.42$, in good agreement (given our time resolution) with the quadrupolar $\ell=2,m=0$ quasinormal mode, $M\omega=0.416 - i0.0765$~\cite{Berti:2005ys}.
}
\label{waveform}
\end{figure}

\noindent {\bf \em GWs from turbulently excited BHs.} 
%
%
The dominant, axially symmetric mode of the GW emitted by the disk is shown in Fig.~\ref{waveform}. The turbulent gas flow excites stochastically the BH modes, which is imprinted in the waveform. We find a frequency of order $M\omega \sim 0.42$ (estimated using the number of cycles within the vertical dashed gray lines in the plot), in very good agreement with the quadrupolar $\ell=2,m=0$ quasinormal mode, $M\omega=0.416 - i0.0765$~\cite{Berti:2005ys}. The ``cross'' polarization $h_\times$ is about an order of magnitude smaller than the ``plus'' $h_+$ reflecting the fact that motion along the radial direction dominates emission. One can expect excitation of higher order modes ($\ell>2$ modes), but these have larger frequencies. To correctly capture these modes we would need prohibitively small (computationally expensive) time steps.

\begin{figure}[t]
\centering
\includegraphics[width = 0.45\textwidth]{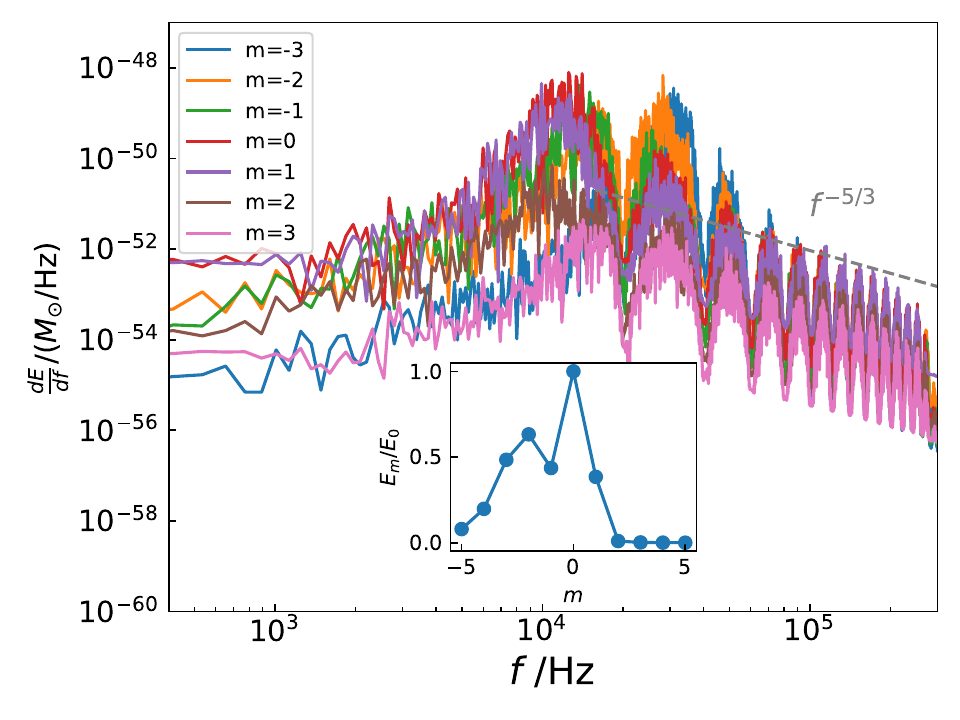}
\caption{GW energy spectrum from accretion disk turbulence with $f_{\mathrm{Edd}}=1$, for several modes. The BH mass is $M=M_{\odot}$. Inset shows the total energy loss of each $m$-mode, $E_m/E_0$.
\label{dEdf} 
}
\end{figure}

Numerical solutions of the Teukolsky equation can generate unphysical signals -- ``junk radiation'' -- as a consequence of arbitrary initial data, which eventually decays in time. We retain the physically meaningful data, for which $t-r_* \gtrsim 300M$, and use it to calculate the energy spectrum of GWs. Fig.~\ref{dEdf} shows the GW energy spectrum for different $m$ modes. To a good accuracy, the spectrum peaks at the lowest quasinormal frequency of the mode in question~\cite{Berti:2005ys,Berti:2009kk}, which is one more piece of evidence that accretion disks are exciting BH ringdown. At higher frequencies, the spectrum decays roughly as a power law. It is tempting to conjecture that the spectrum falls off as $dE/df\sim f^{-5/3}$, corresponding to a Kolmogorov scaling law~\cite{k41a,k41b,k41c}.
The behavior of the energy spectrum at high frequencies (around $10^5$ Hz) closely resembles this Kolmogorov scaling, also an indication of the energy cascade. This suggests the possibility that a dissipative region may exist due to the complex internal dynamics within the accretion disk~\cite{k41a,k41b,k41c}. Our results are then consistent with a turbulent accretion disk stochastically exciting the modes of BHs. The energy spectrum exhibits oscillatory behaviour at high frequencies due to Nyquist sampling. If we employ higher-order interpolation for the source term when solving the Teukolsky equation, the oscillatory behaviour can be mitigated, leading to a power spectrum that scales as $f^{-5/3}$ in the high-frequency band.

The inset of Fig.~\ref{dEdf} shows the relative energy carried in each $m$ mode, $E_m$. The dominant mode is indeed the axial symmetric $m=0$ and the energy emitted quickly decreases to zero (although, as we said, our simulation is unable to resolve very high $m$'s and these might be underestimated). The quick decay of the power spectrum for higher modes is consistent with Ref.~\cite{Wu:2023qeh}, which employs an analytic power series for the turbulent Newtonian potential of the disk.

\noindent {\bf \em Stochastic GW background.} 
\begin{figure}
\centering
\includegraphics[width=0.45\textwidth]{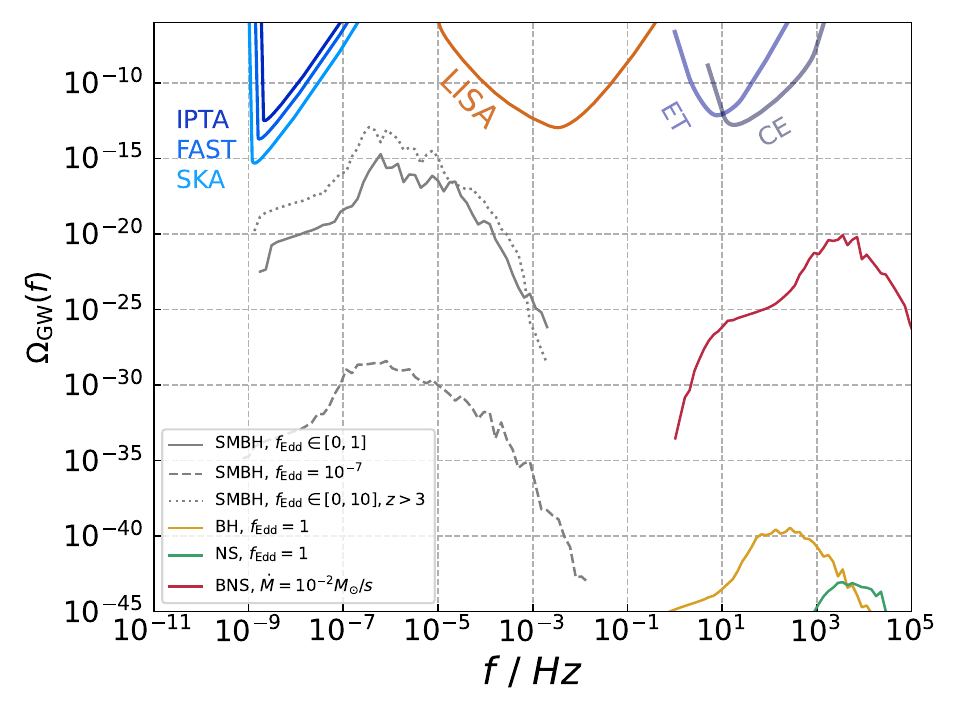}
\caption{\label{SGWB} The SGWB from turbulent accretion disks, compared against the sensitivity curves of GW detectors, including LISA, IPTA, FAST, SKA,  CE and ET.}
\end{figure}
A stochastic GW background (SGWB) is the incoherent superposition of unresolvable GWs. It is characterized by the energy spectrum (energy density per logarithm frequency) normalized by the critical energy:
\begin{equation}
\Omega_{\mathrm{GW}}(f)\equiv\frac{1}{\rho_c}\frac{\mathrm{d} \rhoGW}{\mathrm{d} \ln f}.
\end{equation}
We consider three types of GW-generating disks:

\noindent {\bf A.} Accretion disks around supermassive BHs (SMBH). Sgr A${}^*$ observations suggest an accretion rate $10^{-9}\lesssim\dot{M}\lesssim 10^{-7} (\Msun/\mathrm{yr})$, corresponding to a low ratio $f_{\mathrm{Edd}}\simeq 10^{-7}$ \cite{Aitken_2000,Bower_2003,Marrone:2006vu}. However, a fraction of the SMBHs in the Universe are known to be \textit{actively} accreting, i.e., with accretion rates close to the Eddington limit. For example, observations on J0529–4351 find $f_\mathrm{Edd}\simeq 0.9$ \cite{Wolf:2024lxc}, and recently the James Webb Space Telescope has discovered a population of quasars at $z=6$  begincite{Yang:2021imt}. Therefore, we consider three different populations: one with $f_{\mathrm{Edd}}=10^{-7}$, one with a uniform distribution $f_{\mathrm{Edd}} \in[0,1]$ and {super-Eddington accretion \cite{Suh:2024jbx} with $f_{\mathrm{Edd}} \in[0,10]$ for SMBHs at $z>3$}.
the energy spectrum of SGWB is evaluated as~\cite{Phinney:2001di}
\begin{equation}\label{SGWBeq}
    \Omega_{\mathrm{GW}}(f)=\frac{f}{\rho_c} \int d M d z \frac{d t}{d z} R_{\mathrm{birth}}(z,M) \frac{d E_s}{d f_s} \, ,
\end{equation}
where $R_{\mathrm{birth}}(z,M)$ denotes the BH/NS/SMBH formation rate per comoving volume per mass and ${d E_s}/{d f_s}$ corresponds to the energy spectrum of a single GW event in the source frame. Here ${d t}/{d z}$ is the derivative of the lookback time with respect to the redshift, namely $dt/dz = [H_0 E(z)(1+z)]^{-1}$ where $E(z)=\sqrt{\Omega_r(1+z)^4+\Omega_m(1+z)^3+\Omega_\Lambda}$
%
%
$H_0=67.4~\mathrm{km/s/Mpc}$ the Hubble constant at present \cite{Planck:2018vyg}, $\Omega_r$, $\Omega_m$ and $\Omega_{\Lambda}$ are the density parameters for radiation, matter and dark energy respectively. For $R_{\mathrm{birth}}$, we adopt the SMBH population model at different $z$ in Ref.~\cite{Kelly:2010qc}

\noindent {\bf B.} For an isolated neutron star (NS) (which we take to be described by the Kerr metric~\cite{Frutos-Alfaro:2018gql}) or stellar-origin BH, $f_{\mathrm{Edd}}$ is supposed to be $10^{-2}\lesssim f_{\mathrm{Edd}} \lesssim 1$ or even beyond the Eddington limit \cite{Bachetti:2014qsa,Israel:2016chx,Furst:2016ffn,Brightman:2019nwc,Yoshioka:2024pmo}. We adopt $f_{\mathrm{Edd}}=1$ here.
The SGWB in such a case can also be calculated using Eq.~(\ref{SGWBeq}) where the computation of $R_{\mathrm{birth}}$ for NS and BH can be found in Appendix.~\ref{BHNSbirth}.

\noindent {\bf C.} Finally, we consider the scenario where the central object is formed from the merger of binary neutron stars (BNSs). Such mergers result in a short-lived accretion disk of sub-solar mass~\cite{Kiuchi:2022nin}, corresponding to an accretion rate of $\dot{M}\sim 10^{-2}\Msun/\mathrm{s}$~\cite{Gottlieb:2024dfw}, consistent with  numeric simulations~\cite{Kiuchi:2022nin}.
The corresponding SGWB in this case is evaluated as \cite{Phinney:2001di}
\begin{equation}\label{SGWBeqc}
    \Omega_{\mathrm{GW}}(f)=\frac{f}{\rho_c} \int d M_1d M_2 d z \frac{d t}{d z} \mathcal{R}(z,M_1,M_2)\frac{d E_s}{d f_s},
\end{equation}
and we follow~\cite{LIGOScientific:2016fpe,LIGOScientific:2017zlf} to obtain the merger rate density,
\beq
\mathcal{R}(z,M_1,M_2) &\propto & \int_{t_{\min }}^{t_{\max }} 
R_{\mathrm{birth}}\left(t(z)-t_{d}, M_{1}\right) \nonumber \\
& \times& P_{d}\left(t_{d}\right) P\left(M_{1}\right) P\left(M_{2}\right) d t_{d}\,,\label{Rdensity}
\eeq
with $t(z)$ the age of the Universe at merger. The function $P_d(t_d)\propto 1/t_d$ is the distribution of delay time with $t_{\min }<t_{d}<t_{\max }$. 
Since the local merger rate of BNS is much larger than the BH-NS binaries and binary BHs~\cite{KAGRA:2021duu}, we only consider BNSs in the following computation. For BNSs, $t_{\min }=20$ Myr and $t_{\max }$ is the Hubble time.
The NS mass function has a uniform distribution between $1~\Msun$ to $2~\Msun$ and we normalize the merger rate density so that the local merger rate is given by $\int \mathcal{R}\left(z=0, M_{1}, M_{2}\right) d M_{1} d M_{2}=1000~ \mathrm{yr}^{-1}\,\mathrm{Gpc}^{-3}$ for BNSs~\cite{LIGOScientific:2020kqk,AghaeiAbchouyeh:2023lap}. 

On the other hand, the GRMHD simulation only covers a limited timespan, whereas the age of a real BH far exceeds this value. Therefore, we rescale the amplitude of the GW energy spectrum based on the duration time of the GWs, estimated as follows for all three cases
\begin{equation}
    \Delta t = \min\(t_0-t_{\mathrm{birth}}(z),\frac{M_\mathrm{disk}}{\dot{M}}\),
\end{equation}
where $t_0$ is the age of the universe and  $t_{\mathrm{birth}}(z)$ is the formation time of the central object.

The power-law integrated sensitivity curves~\cite{Thrane:2013oya} of various GW detectors are obtained assuming threshold signal-to-noise ratio (SNR) to be $\mathrm{SNR}=5$ and total observation time of $4$ years for LISA~\cite{LISA:2017pwj}, CE~\cite{LIGOScientific:2016wof} and ET~\cite{Punturo:2010zz}, and $30$ years for IPTA~\cite{Hobbs:2009yy}, FAST~\cite{Nan:2011um} and SKA~\cite{Kramer:2015bea}.
We assume two co-aligned and co-located identical detectors for CE and ET while for LIGO we use the overlap function \cite{Thrane:2013oya}. For IPTA/FAST/SKA we assume the pulsars are uniformly distributed in the sky. The number of pulsars and the timing accuracy can be found in Table~5 of \cite{Kuroda:2015owv}.

The estimated SGWB, including all modes with $|m|\leq 5$ is shown in Fig.~\ref{SGWB}. 
The SGWB generated by the short-lived accretion disks around the remnant of BNSs and the SGWB from isolated NSs/stellar-origin BHs are far below the sensitivity of future ground-based GW detectors.
On the other hand, the SGWB from the SMBH population is peaked around $10^{-6}$ Hz and it could reach  $\Ogw\sim10^{-15}$ for a uniform distribution $f_{\mathrm{Edd}}\in[0,1]$, while it is far below the detection limit if the SMBH population has a small accretion rate $f_{\mathrm{Edd}}=10^{-7}$ as Sgr A${}^*$. 
We also check that the result for $f_{\mathrm{Edd}}=1$ is only slightly larger than $f_{\mathrm{Edd}}\in[0,1]$ since the GW emission is dominated by SMBHs with large accretion rate. 

Although the GRMHD simulations are traditionally employed for modeling accretion disks around SMBHs, we extend their application to systems involving stellar-origin BHs and NSs. It is worth noting that the direct application of the GRMHD simulations to stellar mass objects may introduce uncertainties due to different physical conditions, such as differences in magnetic field configurations.

\noindent {\bf \em Discussion.} 
In this Letter, we explore the contribution of turbulence in accretion disks around BHs and NSs to the emission of GWs, in particular to stochastic backgrounds. Feeding GRMHD simulations as a source term to the Teukolsky equation, we have quantified the GW energy spectrum from these systems. The waveform and energy spectrum analysis indicate that turbulent flows in accretion disks stochastically excite the quasinormal modes of the BHs.
The energy spectrum shows a power-law decay, which scales as the expected Kolmogorov decay $dE/df\sim f^{-5/3}$ for a turbulent system.
Our results indicate that the SGWB from accretion disks around NSs, stellar-origin BHs and the remnants from the merger of BNSs are undetectable due to their low amplitudes. 
Note that Eq.~(\ref{SGWBeqc}) implies that we assume all the BNS merger lead to short-lived accretion disks. However, the SGWB is still far below the sensitivity curves, indicating case {\bf C.} is of low interest despite of their large accretion rates.
We neglect differences between QNMs of NSs and those of BHs~\cite{schutz2008asteroseismology}. However, as the SGWB from NSs and BNSs are far below the sensitivity of future detectors, this simplification will most likely not affect our overall conclusions.
On the other hand, accretion disks around SMBHs could generate an SGWB at $\Ogw\sim 10^{-15}$ in the microHertz frequencies.
{We also explored the astrophysical implications of super-Eddington accreting systems at $z>3$. The resulting $\Omega_{\mathrm{GW}}\sim10^{-14}$ might be an interesting prospect for future GW detectors in this frequency band.
}
Additionally, our findings motivate multi-messenger studies that link the GW signals from accretion disk-induced mode excitation to electromagnetic observations of accretion disks around compact objects. This could provide a unique window into the dynamics of accretion flows and the environments of SMBHs.


\noindent {\bf \em Acknowledgment.} 
%
We acknowledge the support by VILLUM Foundation (grant no. VIL37766) and the DNRF Chair program (grant no. DNRF162) by the Danish National Research Foundation.
V.~C.\ is a Villum Investigator and a DNRF Chair.  
C.~Y.\ and V.~C. acknowledge financial support provided under the European Union’s H2020 ERC Advanced Grant “Black holes: gravitational engines of discovery” grant agreement no. Gravitas–101052587. 
Views and opinions expressed are however those of the author only and do not necessarily reflect those of the European Union or the European Research Council. Neither the European Union nor the granting authority can be held responsible for them.
Z.~Y.\ is supported by a UK Research and Innovation (UKRI) Stephen Hawking Fellowship.
This project has received funding from the European Union's Horizon 2020 research and innovation programme under the Marie Sklodowska-Curie grant agreement No 101007855 and No 101131233.

\bibliography{./ref}
\clearpage
\onecolumngrid
\appendix

\section{The Teukolsky equation}\label{source}
The Teukolsky equation expressed in {\it Boyer-Lindquist coordinates} reads
\beq
&\Delta^{2}\!\!\!\!&\partial_r\left(\Delta^{-1} \partial_r \Psi\right)-\left[\frac{\left(r^2+a^2\right)^2}{\Delta}-a^2 \sin ^2 \theta\right] \partial_{t t} \Psi +\!\!4\left[r-\frac{M\left(r^2-a^2\right)}{\Delta}+i a \cos \theta\right] \partial_t \Psi +\frac{1}{\sin \theta} \partial_\theta\left(\sin \theta \partial_\theta \Psi\right)\nonumber \\
&+&\!\!\!\! {\left[\frac{1}{\sin ^2 \theta}-\frac{a^2}{\Delta}\right] \partial_{\phi \phi} \Psi} -4\left[\frac{a(r-M)}{\Delta}+\frac{i \cos \theta}{\sin ^2 \theta}\right] \partial_\phi \Psi  \!- \!\left(4 \cot ^2 \theta+2\right) \Psi \!- \!\frac{4 M a r}{\Delta} \partial_{t \phi} \Psi =-4 \pi\left(r^2\!\!+\!\!a^2 \cos ^2 \theta\right) T,~\label{Teukolsky}
\eeq
where $a=M\chi$ is the BH angular momentum,  $\Delta=r^2-2 M r+a^2$, and $T$ is a source term computed from the stress-energy tensor. In Boyer-Lindquist coordinates the tetrad vectors are defined as:
\be
n^{\mu} =\left(\frac{\rho\bar{\rho}\left(r^2+a^2\right)}{2},-\frac{\Delta \rho\bar{\rho}}{2}, 0, \frac{a \rho\bar{\rho}}{2}\right)\,,\,
\bar{m}^{\mu} = \frac{(-i a \sin \theta, 0,1,-i \csc \theta)}{\sqrt{2}(r-i a \cos \theta)}\,,
\ee
where $\rho^{-1}=-(r-i a \cos \theta)$ and a bar represents the complex conjugate. The Newman-Penrose operators are:
\be
\tilde{\Delta}= {n^\mu}\frac{\mathrm{d}}{\mathrm{d}x^\mu}\,,\,\tilde{\delta}=\bar{m}^\mu\frac{\mathrm{d}}{\mathrm{d}x^\mu}\,.
\ee

The source term of the Teukolsky equation can be expressed in a generic form such that
\begin{equation}
\begin{aligned}
T= & 2 (r-ia\cos\theta)^4 T_{4}, \\
T_{4}= & (\tilde{\Delta}+3 \gamma-\bar{\gamma}+4 \mu+\bar{\mu})(\tilde{\Delta}+2 \gamma-2 \bar{\gamma}+\bar{\mu}) T_{\bar{m} \bar{m}} \\
& -(\tilde{\Delta}+3 \gamma-\bar{\gamma}+4 \mu+\bar{\mu})(\bar{\delta}-2 \bar{\tau}+2 \alpha) T_{n \bar{m}} \\
& +(\bar{\delta}-\bar{\tau}+\bar{\beta}+3 \alpha+4 \pi)(\bar{\delta}-\bar{\tau}+2 \bar{\beta}+2 \alpha) T_{n n} \\
& -(\bar{\delta}-\bar{\tau}+\bar{\beta}+4 \pi)(\tilde{\Delta}+2 \gamma+2 \bar{\mu}) T_{n \bar{m}}\,.
\end{aligned}
\end{equation}
The projected stress-energy tensor is defined as $T_{n n}=n^\mu n^\nu T_{\mu \nu}$, $T_{n \bar{m}}=n^\mu \bar{m}^\nu T_{\mu \nu}$ and $T_{\bar{m} \bar{m}}=\bar{m}^\mu \bar{m}^\nu T_{\mu \nu}$.
The Newmann-Penrose scalars are given by (see e.g., \cite{Loutrel:2020wbw}):
\begin{equation}
\begin{aligned}
& \rho=-\frac{1}{\bar{\Gamma}}, \quad \beta=\frac{\cot \theta}{2^{3 / 2} \Gamma}, \quad \pi=\frac{i a \sin \theta}{2^{1 / 2} \bar{\Gamma}^2}, \\
& \tau=-\frac{i a \sin \theta}{2^{1 / 2} \Gamma \bar{\Gamma}}, \quad \mu=-\frac{\Delta}{2 \Gamma \bar{\Gamma}^2}, \quad \gamma=\mu+\frac{r-M}{2 \Gamma \bar{\Gamma}}, \\
& \alpha=\pi-\bar{\beta}, \quad \Psi_2=-\frac{M}{\bar{\Gamma}^3},
\end{aligned}
\end{equation}
with $\Gamma=r+i a \cos \theta$. 

\texttt{BHAC} uses Kerr-schild coordinates, with which the tetrad vectors become:
\begin{equation}
    n^{\mu}=\left[\frac{1}{2 \Sigma},-\frac{1}{2 \Sigma}, 0,0\right],
\end{equation}
where $\Sigma=r^2+a^2\cos^2\theta$ while $\bar{m}$ remains unchanged. The Newmann-Penrose scalars become:
\be
\epsilon=r-M\,,\,\gamma=\mu=-\frac{1}{2} \frac{r+i a \cos \theta}{\Sigma^2}\,,\, \rho=-(r+i a \cos \theta) \frac{\triangle}{\Sigma}\,.
\ee
The rest of the quantities remain unchanged. Finally, the Newman-Penrose operator becomes:
\begin{equation}
    \tilde{\Delta} = n^\mu \frac{\mathrm{d}}{\mathrm{d} x ^\mu} = \frac{1}{2\Sigma}\(\frac{\mathrm{d}}{\mathrm{d} t}-\frac{\mathrm{d}}{\mathrm{d} r}\),
\end{equation}
while $\tilde{\delta}$ remains unchanged.

\section{Black Hole/Neutron Star Formation Rate}\label{BHNSbirth}
We follow Ref.~\cite{Dvorkin:2016wac} to compute the BH/NS formation rate which takes the form
\begin{equation}\label{rbirth}
    R_{\mathrm{birth}} = \int \psi[t - \tau(M_*)]\phi(M_*)\delta(M_* - g_{\mathrm{rem}}^{-1}(M, z)) \mathrm{d}M_* , 
\end{equation}
where $M_*$ is the progenitor star mass and $\psi(z)$ is the star formation rate which takes the form
\begin{equation}
    \psi(z) = \nu \frac{a \exp[b(z - z_m)]}{a - b + b \exp[a(z - z_m)]},
\end{equation}
and the parameters are given by $\nu = 0.178 \, M_\odot \, \text{yr}^{-1} \, \text{Mpc}^{-3},  z_m = 2.00,  a = 2.37,  b = 1.80$ \cite{Vangioni:2014axa}.
The lifetime of the progenitor star in Eq.~(\ref{rbirth}), $\tau(M_*),$ can be computed as \cite{Schaerer:2001jc}
\begin{equation}
    \log_{10}\tau(M_*) = 9.785-3.759x+1.413x^2-0.186x^3,
\end{equation}
with $x= \log_{10}(M/M_{\odot})$. 
The initial mass function is given by $\phi(M_*)\propto M_*^{-2.35}$ for BHs \cite{Salpeter:1955it} and we consider a uniform distribution between $1M_{\odot}$ and $2M_{\odot}$ for NSs. The mass of a BH/NS remnant is related to the mass of the progenitor star by $M = g_{\mathrm{rem}}(M_*, z)$ and can be evaluated as \cite{Fryer:2011cx}
\begin{equation}
M=
\begin{cases} 
1.28, & M_* < 11 \, M_\odot, \\
1.1 + 0.2 e^{(M_* - 11.0)/4.0} - (2.0 + Z(z)/Z_{\odot}) e^{0.4 (M_* - 26.0)}, & 11 \, M_\odot \leq M_* < 30 \, M_\odot, \\
\min\left(33.35 + (4.75 + 1.25 Z(z)/Z_{\odot})(M_* - 34), \, M_* - \sqrt{Z(z)/Z_{\odot}}(1.3 M_* - 18.35)\right), & 30 \, M_\odot \leq M_* < 50 \, M_\odot, \\
1.8 + 0.04 \times (90 - M_*), & 50 \, M_\odot \leq M_* < 90 \, M_\odot, \\
1.8 + \log_{10}(M_* - 89), & M_* \geq 90 \, M_\odot,
\end{cases}
\end{equation}
where $Z_{\odot}=0.0196$ is metallicity of the Sun \cite{Vagnozzi:2017wge}. The metallicity of the progenitor star as a function of the redshift is given by 
\cite{Belczynski:2016obo}
\begin{equation}
\log_{10}Z(z) = 0.5 + \log_{10} \left( \frac{0.01387}{\rho_b} \int_{z}^{20} \frac{97.8 \times 10^{10}  \psi(z')}{H_0 E(z') (1 + z')} \, dz' \right),
\end{equation}
where $\rho_b=6.1\times 10^9M_{\odot}\mathrm{Mpc}^{-3}$ and $E(z) = \sqrt{\Omega_r(1+z)^4+\Omega_m(1+z)^3+\Omega_\Lambda}$.

\end{document}